\input{aipcheck}

\documentclass[
    ,final            
  ]
  {aipproc}

\layoutstyle{6x9}

\begin{document}

\title{In search of the perfect fluid}


\author{Thomas Sch\"afer}{
  address={Department of Physics, North Carolina State University, 
           Raleigh, NC 27695}
}

\begin{abstract}
Shear viscosity measures the amount of internal friction in a 
simple fluid. In kinetic theory shear viscosity is related to momentum 
transport by quasi-particles, and the uncertainty relation implies 
that the ratio of shear viscosity $\eta$ to entropy density $s$ is 
bounded by a constant multiplied by $\hbar/k_B$, where $\hbar$ is 
Planck's constant and $k_B$ is Boltzmann's constant. A specific 
bound has been proposed on the basis of string theory. In a large 
class of theories that can be studied using string theory methods
the constant is $1/(4\pi)$. Experiments at RHIC indicate that $\eta/s$ 
of the quark gluon plasma is close to this prediction. We will refer 
to a fluid that saturates the string theory bound as a perfect fluid. 
In this contribution we summarizes the theoretical and experimental 
information on the fluidity of the main classes of strongly 
interacting quantum fluids. 
\end{abstract}

\maketitle


\section{Introduction}

Experiments at the relativistic heavy ion collider indicate
that the quark gluon plasma behaves more like a strongly 
coupled liquid than like a weakly coupled plasma. The most
dramatic manifestation of this liquid-like behavior is the 
observation of almost ideal hydrodynamic flow in heavy ion 
collisions \cite{Adler:2003kt}. This experimental result 
is in contrast to earlier experiments at lower energies, 
and to theoretical expectations based on weak-coupling 
QCD. 

 Corrections to ideal hydrodynamics are governed by 
dissipative terms. The magnitude of these terms is determined
by transport coefficients, in particular shear viscosity, 
bulk viscosity, and thermal conductivity. We will see that
for very good fluids the main source of dissipation is shear
viscosity. Shear viscosity can be defined in terms of the 
friction force $F$ per unit area $A$ created by a shear flow 
with transverse flow gradient $\nabla_{\! y} v_x$,
\begin{equation}
\label{eta_fric}
 \frac{F}{A}=\eta\, \nabla_{\! y} v_x\, .
\end{equation}
In a weakly coupled gas of quasi-particles the shear 
viscosity can be estimated as 
\begin{equation}
\label{eta_mfp}
\eta = \frac{1}{3}\,n p l_{\it mfp}\, ,
\end{equation}
where $n$ is the density, $p$ is the average momentum of the
particles, and $l_{\it mfp}$ is the mean free path.  The mean free
path can be written as $l_{\it mfp}=1/(n\sigma)$ where $\sigma$
is a suitable transport cross section. 

 The shear viscosity of a good fluid is small. But just how
small can $\eta$ get?  Danielewicz and Gyulassy pointed out 
that the Heisenberg uncertainty relation imposes a bound 
on the product of the average momentum and the mean free
path, $p l_{\it mfp}\geq \hbar$, and concluded that $\eta/n
\geq \hbar$ \cite{Danielewicz:1984ww}. For relativistic fluids
it is more natural to normalize $\eta$ to the entropy density 
rather than the particle density. Using $s\sim k_B n$ we 
conclude that $\eta/s\geq \hbar/k_B$. 

 This is not a precise statement: The kinetic estimate in 
equ.~(\ref{eta_mfp}) is not valid if the mean free path 
is on the order of the mean momentum. An important recent 
development is the discovery that one can compute the 
strong coupling limit of the ratio $\eta/s$ in certain
extensions of QCD using methods borrowed from string theory. 
Policastro et al.~ showed that in ${\cal N}=4$ supersymmetric
QCD the strong coupling limit of $\eta/s$ is equal to $\hbar/
(4\pi k_B)$ \cite{Policastro:2001yc}, and it was later conjectured
that $\eta/s \geq\hbar/(4\pi k_B)$ is a universal lower 
bound, valid for all fluids \cite{Kovtun:2004de}.

\section{Perfect Fluidity}

 We will refer to a fluid that saturates the proposed bound
as a ``perfect fluid''. A perfect fluid is not only characterized 
by very small dissipation, but also by the fact that hydrodynamics
has the largest possible domain of validity. In a typical fluid
hydrodynamics is an effective description of low frequency, 
long wavelength fluctuations. In a perfect fluid hydrodynamics 
is valid down to scales as small as the inter-particle spacing. 

 Are there any perfect or nearly perfect fluids in nature? A
perfect fluid has to be a quantum fluid (because $\eta$ is
on the order of $\hbar n$), and it has to be strongly interacting
(because in a weakly interacting system the mean free path is 
large). The main examples of strongly interacting quantum fluids
in nature are i) liquid $^4$He, a strongly coupled Bose fluid, 
ii) dilute Fermi gases at unitarity (systems in which the 
scattering length was tuned to infinity using a Feshbach 
resonance), iii) strongly coupled plasmas, in particular the 
quark gluon plasma. 

\begin{figure}
\includegraphics[width=7.2cm]{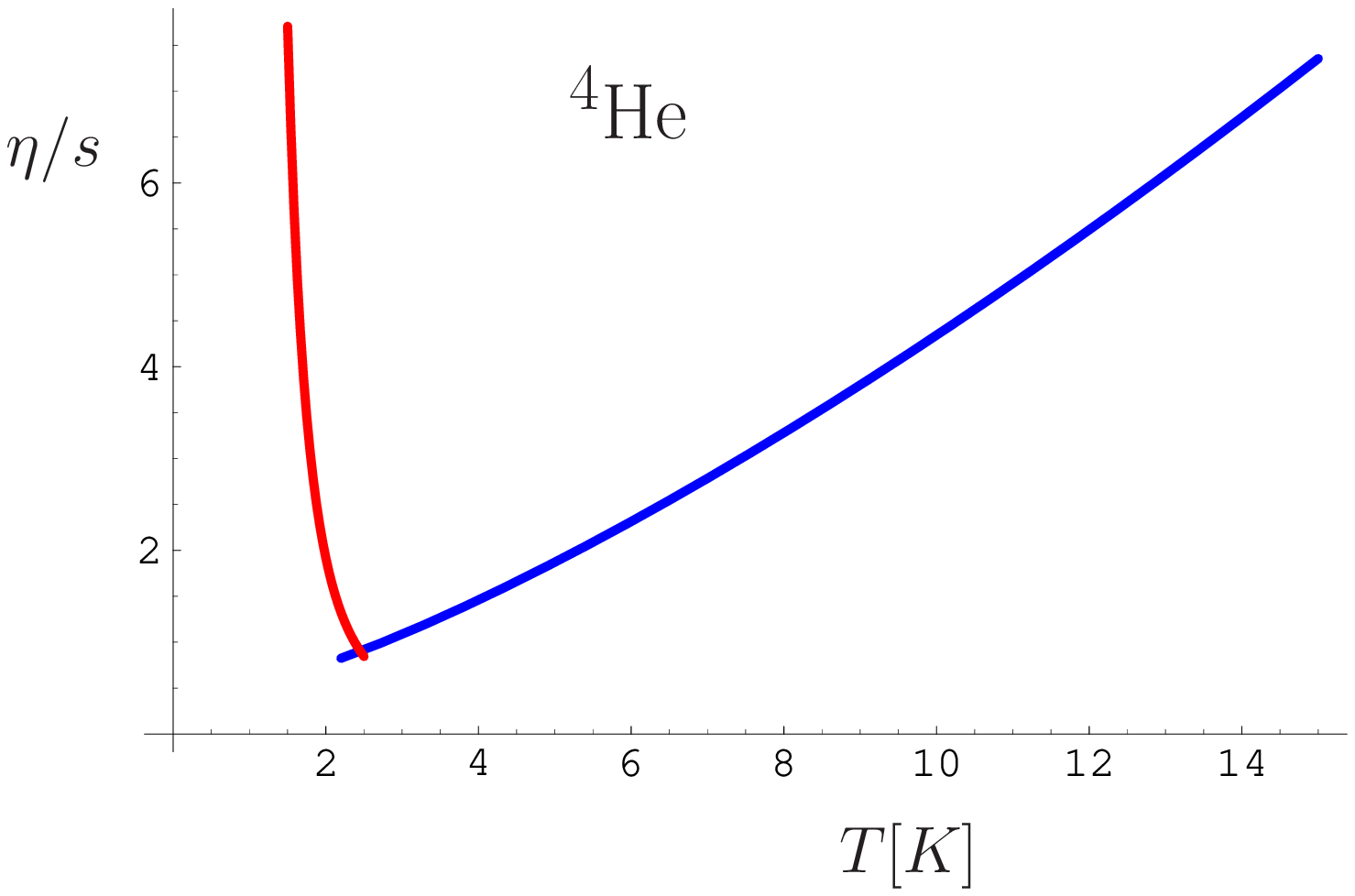}
\includegraphics[width=7.2cm]{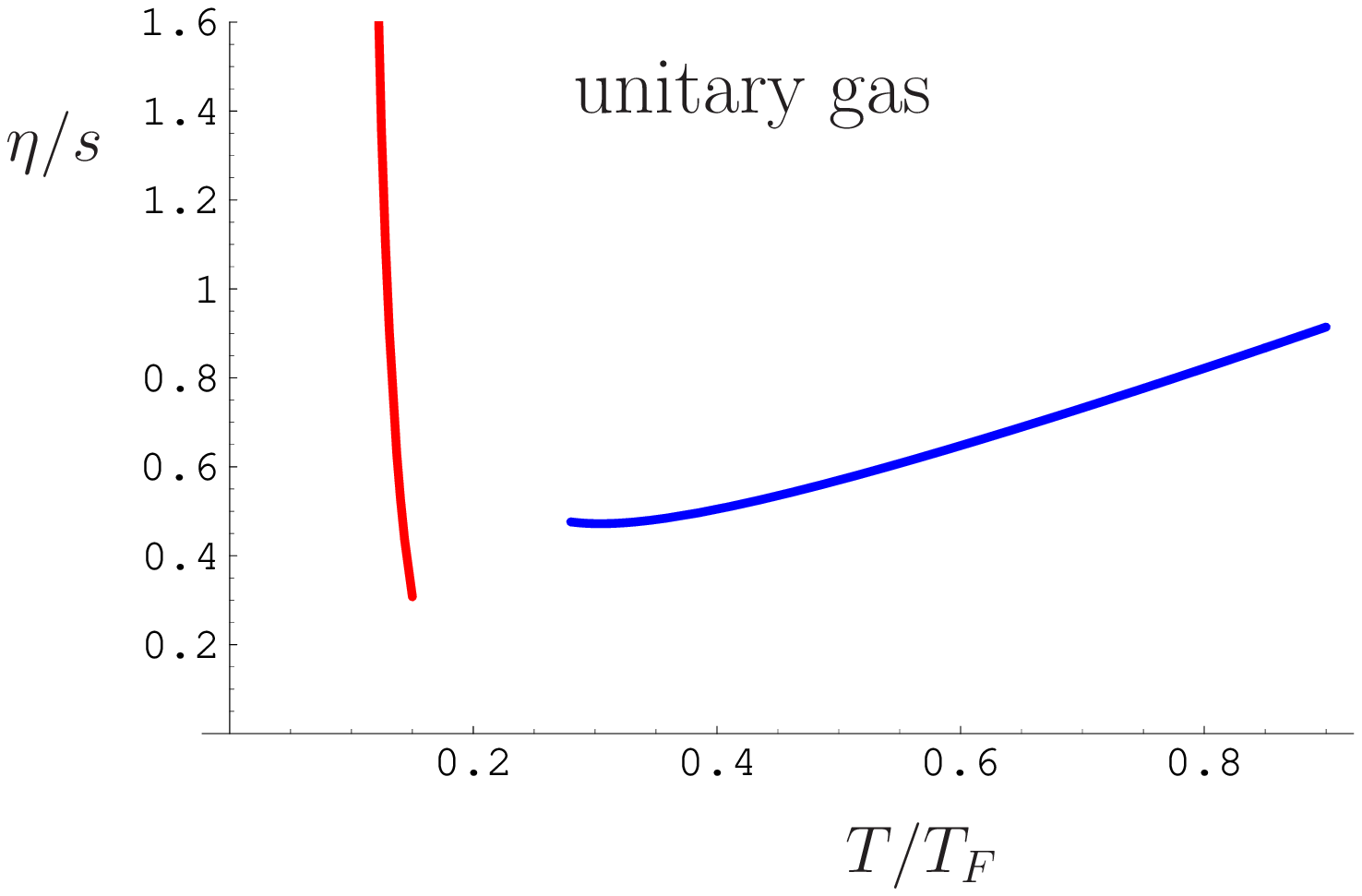}
\caption{Kinetic theory prediction for the ratio $\eta/s$
in units of $\hbar/k_B$ as a function of temperature in liquid 
$^4$He (left panel) and for a dilute Fermi gas at unitarity 
(right panel). The temperature of helium is given in Kelvin. 
For the dilute Fermi gas $T$ is given in units of the Fermi 
temperature $k_BT_F=(3\pi^2 n)^{2/3}/(2m)$ where $m$ is the 
mass of the atoms and $n$ is the density. The red curves show the low 
temperature expansion, and the blue curve is the high temperature 
expansion. }
\end{figure}

\section{Transport Theory}

 In this section we summarize theoretical information about 
the transport properties of these quantum fluids. We will 
concentrate on results based on kinetic theory. For more 
details and for a discussion of other approaches (holography, 
linear response theory, etc) we refer the reader to our 
recent review \cite{Schafer:2009dj}.

 Kinetic theory applies whenever the fluid can be described
in terms of quasi-particles. For many fluids this is the case
in both the high and the low temperature limit. Typically, at 
high temperature the quasi-particles are the ``fundamental''
degrees of freedom (atoms in the case of helium and the dilute 
Fermi gas, quarks and gluons in the case of QCD) whereas the low
temperature degrees of freedom are composite (phonons and rotons
or just phonons in the atomic systems, and pions in the case 
of QCD). In the regime in which kinetic theory applies the 
ratio $\eta/s$ is always parametrically large. This leads to 
a characteristic ``concave'' temperature dependence of $\eta/s$
\cite{Csernai:2006zz}. Kinetic theory is useful in constraining 
the location of the viscosity minimum, but it cannot reliably 
predict the minimum value of $\eta/s$. 

 Kinetic theory results are summarized in Figures 1 and 2. 
We observe that all three fluids are expected to exhibit 
viscosity minima with $(\eta/s)_{\it min}<\hbar/k_B$. The 
minima occur in the vicinity of the phase transition, $T\sim 
2.2$ K in the case of helium, $T\sim 0.15 T_F$ for the 
dilute Fermi gas, and $T\sim (150-180)$ MeV in QCD.

\begin{figure}
\includegraphics[width=7.2cm]{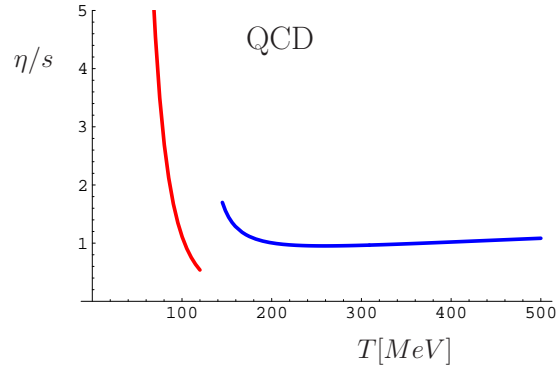}
\caption{Kinetic theory prediction for $\eta/s$ in QCD.The red 
curve shows the low temperature expansion, and the blue curve 
is the results of the high temperature expansion. }
\end{figure}

\section{Experimental Situation}

 Finally, we summarized the experimental situation for 
the the three quantum fluids discussed in this contribution. 

\begin{itemize}
\item Liquid helium has been studied for many years and its 
shear viscosity is well determined. The minimum value of 
$\eta/s$ is about 0.8 $\hbar/k_B$ and is attained near the 
endpoint of the liquid gas phase transition. The ratio $\eta/n$ 
has a minimum closer to the lambda transition. Even though 
$\eta/s<\hbar/k_B$ the transport properties of liquid helium 
can be quantitatively understood using kinetic theory
\cite{Khalatnikov:1965}.

\item Strongly interacting cold atomic Fermi gases were first 
created in the laboratory in 1999. These systems are interesting 
because the interaction between the atoms can be controlled,
and a large set of hydrodynamic flows (collective oscillations,
elliptic flow, rotating systems) can be studied. Current
experiments involve $10^5-10^6$ atoms, and the range of 
temperatures and interaction strengths over which hydrodynamic
behavior can be observed is not large. 
An analysis of the damping of collective oscillations gives 
$\eta/s\sim 0.5$ \cite{Schafer:2007pr,Turlapov:2007}. Even 
smaller values of $\eta/s$ are indicated by recent data on 
the expansion of rotating clouds \cite{Clancy:2007}.

\item The quark gluon plasma has been studied in heavy ion
collisions at a number of facilities, AGS (Brookhaven), SPS
(CERN), RHIC (Brookhaven). Almost ideal hydrodynamic behavior
was observed for the first time in 200 GeV per nucleon (in the 
center of mass) Au on Au collisions at RHIC. These experiments
are difficult to analyze - the initial state is very far from
equilibrium and not completely understood, final state interactions 
are important, and the size and lifetime of the system are not
very large. Important progress has nevertheless been made in 
extracting constraints on the transport properties of the 
quark gluon plasma. A conservative bound is $\eta/s<0.4$, but 
the the value of $\eta/s$ that provides the best fit to the 
data is smaller, $\eta/s\sim 0.1$ \cite{Heinz:2009xj}.

\end{itemize}

\bibliographystyle{aipproc}   

\end{document}